\PassOptionsToPackage{hyphens}{url}
\documentclass[letterpaper,twocolumn,10pt]{article}
\usepackage{usenix2019_v3}

\usepackage{url}
\usepackage{breakurl}
\usepackage{adjustbox}
\usepackage{graphicx}
\usepackage[utf8]{inputenc}
\usepackage{enumitem}
\usepackage{multirow}
\let\oldbibitem\bibitem
\def\bibitem{\vfill\oldbibitem}
\providecommand{\myparab}[1]{\smallskip\noindent\textbf{#1}\hspace{0.6em}}
\newcommand{\sectionref}[1]{$\S$\ref{#1}}
\begin{document}
%-----------------------------------------------------------------------------

%don't want date printed
\date{}

% potential reframing title to focus on the contribution of I2P measurement
\title{Measuring I2P Censorship at a Global Scale}

%for single author (just remove % characters)
\author{
% {\rm Anonymous Author(s)}
{\rm Nguyen Phong Hoang}\\
Stony Brook University
\and
{\rm Sadie Doreen}\\
The Invisible Internet Project
\and
{\rm Michalis Polychronakis}\\
Stony Brook University
} % end author

\maketitle

\begin{abstract}

The prevalence of Internet censorship has prompted the creation
of several measurement platforms
for monitoring filtering activities. An important challenge faced by these
platforms revolves around the trade-off
between depth of measurement and breadth of coverage.
In this paper, we present an opportunistic censorship
measurement infrastructure built on top of a network of distributed VPN
servers run by volunteers, which we used to measure the extent to which the
I2P anonymity network is blocked around the world.
This infrastructure provides us with not
only numerous and geographically diverse vantage points,
but also the ability to conduct
in-depth measurements across all levels of the network stack.
Using this infrastructure, we measured at a global scale
the availability of four different I2P services: the official
homepage, its mirror site, reseed servers, and active relays in the network.
Within a period of one month, we conducted a total of 54K measurements from
1.7K network locations in 164 countries. With different techniques for detecting
domain name blocking, network packet injection, and block pages, we 
discovered I2P censorship in five countries: China, Iran, Oman, Qatar, and
Kuwait. Finally, we conclude by discussing potential approaches to
circumvent censorship on I2P.

\end{abstract}

\section{Introduction}

Several platforms have been built to measure Internet censorship at a large
scale, including the OpenNet Initiative~\cite{gill.2015.worldwide},
ICLab~\cite{iclab_SP20}, Open Observatory of Network Interference
(OONI)~\cite{ooni-paper}, Quack~\cite{VanderSloot:2018:Quack},
Iris~\cite{Pearce:2017:Iris}, and Satellite~\cite{Will:2016:Satellite}.
A common challenge faced by these platforms is the trade-off between depth
of measurement and breadth of coverage.

In this paper, we present a complementary measurement infrastructure
that can be used to address the above issue. The infrastructure is built on
top of a network of distributed VPN servers operated by volunteers around the
world. While providing access to many residential network locations, thus
addressing the coverage challenge, these servers also offer the
required flexibility for
conducting fine-grained measurements on demand.
We demonstrate these benefits by conducting an in-depth investigation
of the extent to which the I2P
(invisible Internet project) anonymity network is blocked across different
countries.

Due to the prevalence of Internet censorship and online surveillance in recent
years~\cite{FreedomHouse2018, Aase:FOCI12, RussiaTelegram}, many pro-privacy
and censorship circumvention tools, such as proxy servers, virtual private
networks (VPN), and anonymity networks have been developed. Among these tools,
Tor~\cite{Dingledine:Usec13} (based on onion
routing~\cite{syverson96OnionRouting, syverson97anonymous}) and
I2P~\cite{zzz:petcon09} (based on garlic routing~\cite{Dingledine2000,
Dingledine2001, freedman2000design}) are widely used by privacy-conscious and
censored users, as they provide a higher level of privacy and
anonymity~\cite{hoang:icact14}.

In response, censors often hinder access to
these services to prevent their use~\cite{Winter:foci12,
Ensafi:imc15, Arun:foci18}. Therefore, continuous measurements are essential
to understand the extent of filtering and
help in restoring connectivity to these networks for end
users~\cite{Nishihata:IIC2013}. While many works have studied censorship on
Tor~\cite{Winter:foci12, Ensafi:imc15, Arun:foci18}
(OONI~\cite{ooni-paper} even has a dedicated
module to test connectivity to the Tor network),
none have comprehensively examined the
blocking status of I2P. To fill this gap, in this work we investigate the
accessibility of I2P
using the proposed VPN-based measurement infrastructure.

By conducting 54K measurements from 1.7K vantage points in 164 countries
during a one-month period, we found that China hindered access to I2P by
poisoning DNS resolutions of the I2P homepage and three reseed servers.
SNI-based blocking was detected in Oman and Qatar when accessing the I2P
homepage over HTTPS. TCP packet injection was detected in Iran, Oman, Qatar,
and Kuwait when visiting the mirror site via HTTP. Explicit block pages
were discovered when visiting the mirror site from Oman, Qatar, and Kuwait.
Based on these findings, we conclude by discussing potential approaches for
improving I2P's resistance to censorship.

\section{Background}
\label{sec:background}

In this section, we review the VPN Gate ecosystem~\cite{VPNGate} and the basic
operation of the I2P anonymity network~\cite{zzz:petcon09}.

\subsection{VPN Gate}
\label{sec:vpngate}
VPN Gate is an academic project developed at the University of Tsukuba,
Japan~\cite{Nobori2014}. Its core component is a network of distributed VPN
vantage points hosted by volunteers from around the world. Unlike
commercial VPNs, these VPN vantage points are operated by Internet users who
are willing to share their home connection, with the primary goal to provide
other users with access to the Internet. Volunteers use a software
package called SoftEther VPN~\cite{SoftEther} to turn their personal computer
into a VPN server. Other users can then establish VPN connections to these 
servers using the client component of the same VPN software package.

\myparab{Advantages.} Since VPN Gate's vantage points (VGVPs) are organized
and operated by volunteers, they provide three essential benefits
that make them a potential resource for measuring censorship at a global scale.
First, VGVPs are often located in residential networks, and can help
to observe filtering policies which may not be observed when measuring from
non-residential networks (e.g., data centers). Second, VGVPs provide access to
many network locations that are difficult to obtain through commercial VPNs. Our
results (\sectionref{sec:tcp_injection}) indeed show that having access
to several network locations is important for observing different blocking
policies, even within the same country.

Finally, unlike commercial VPNs that often monetize their services
by injecting advertisements~\cite{Ikram:IMC16, Khan:2018:VPN} or even ``lying''
about their geographical location~\cite{Weinberg:2018:CPL:Proxy}, VGVPs
managed by individual operators are unlikely to carry out such illicit
practices---though this possibility cannot be excluded,
as rogue network relays have been found in Tor~\cite{torsniff}.
Even if a VGVP is malicious, the chance
that it is selected for our measurements is small, given the
thousands of available VGVPs. We actually actively looked for and 
did not observe any malicious
JavaScript or ad injection in our measurements.

\myparab{Limitations.} As VGVPs are run by individuals on their
personal computers, they cannot guarantee continuous uptime. We can therefore
only use them to conduct measurements when they are online. Another drawback
of using VGVPs is that their availability is susceptible to blocking based on
protocol signatures. Local Internet authorities can prevent VGVPs from
functioning by filtering the VPN protocols supported by VPN Gate, including
L2TP/IPsec, OpenVPN, MS-SSTP, and SSL-VPN (of which OpenVPN is the most
prevalent protocol). In that case, we would not have access to VGVPs in locations
where such filtering policies are applied. VPN Gate
mitigates this problem by allowing VGVPs to run on random ports, instead of
the default ports of the aforementioned VPN protocols.

\subsection{The Invisible Internet Project}
\label{sec:i2p}

\begin{figure}[t]
\centering
\includegraphics[width=\columnwidth]{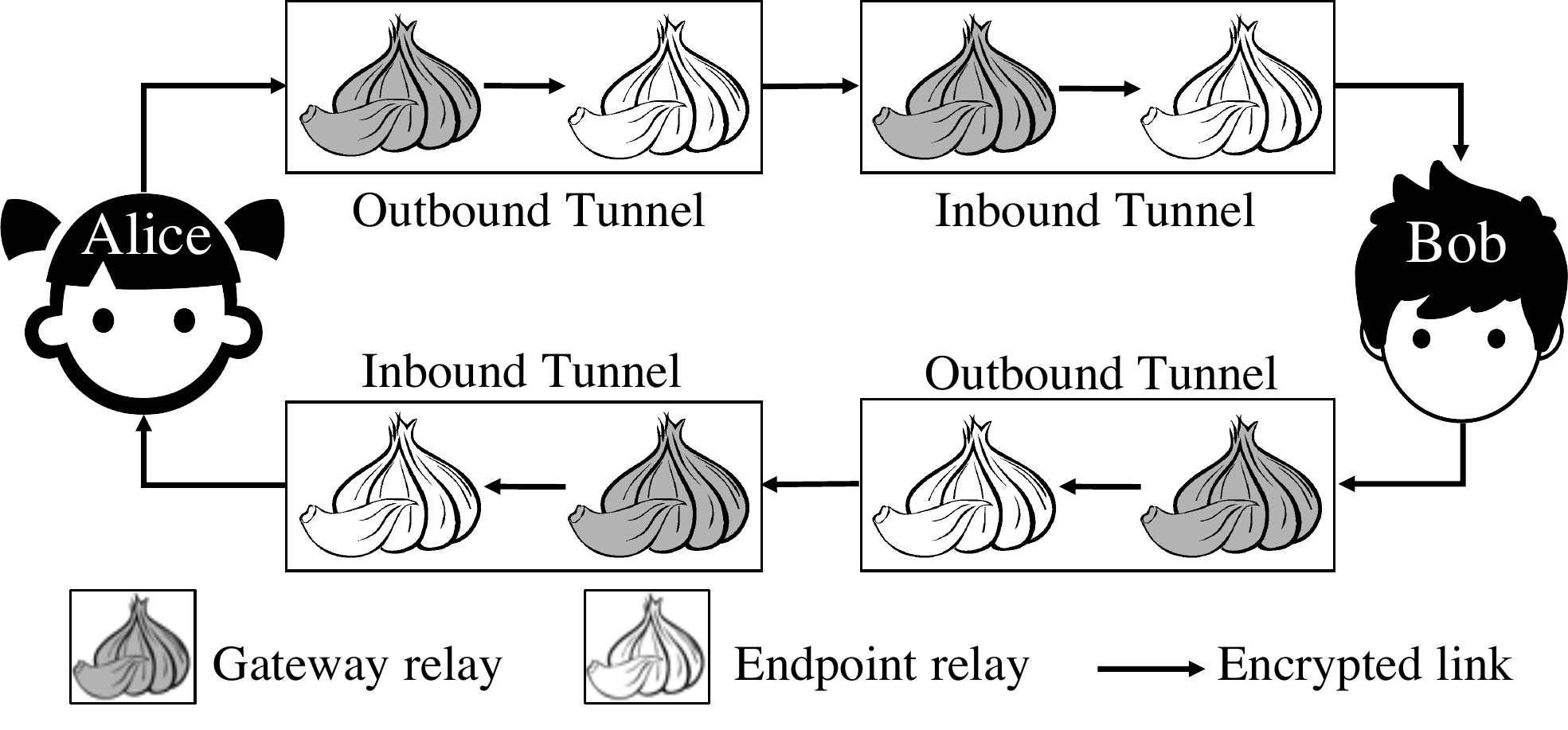}
\vspace{-7mm}
\caption{I2P routing mechanism~\cite{i2parchitecture}.}
\vspace{-5mm}
\label{fig:i2p_routing}
\end{figure}

I2P is a message-oriented anonymous overlay network comprising of relays (also
referred to as nodes, routers, or peers) that run the I2P router software to
communicate with each other. I2P messages are routed through two types of
unidirectional tunnels: inbound and outbound. In the example of
Figure~\ref{fig:i2p_routing}, each tunnel is illustrated with two hops for
simplicity. For a higher level of anonymity, these tunnels can be configured
to have up to seven hops.

To communicate with Bob, Alice sends out messages on her outbound tunnels
towards the inbound tunnels of Bob. Messages from Bob are sent to Alice in the same
way. Alice and Bob learn each other's gateway relay address by querying
a network database. The anonymity of both Alice and Bob is preserved since
they only know the gateway address, but not the actual address of each other.
Note that gateways of inbound tunnels are published, while gateways of
outbound tunnels are only known by the relay using them.

The I2P network database (netDb) originates from the Kademlia distributed hash
table~\cite{Maymounkov2002} and plays a vital role in the network, as it is
used by relays to look up information of other relays. A newly joined relay
learns a portion of the netDb via a bootstrapping process, fetching other
relays' information from a group of special relays called \emph{reseed
servers}. Any I2P relay, when communicating with its intended destination,
can also route traffic for other relays. In Figure~\ref{fig:i2p_routing}, the
hops that are selected to form the tunnels are also actual I2P users. While
routing messages for Alice and Bob, these hops can also communicate with their
intended destinations.

Although Tor and I2P share similar properties, there are some operational
differences. Tor traffic is transmitted over TCP, while I2P traffic can be
transmitted over either TCP or UDP. Tor has a centralized design with a set of
trusted directory authorities keeping track of the network. In contrast, I2P
is designed to be a completely decentralized network, with no trusted entity
having a complete view of the network.

There are 6.3K Tor routers serving an average of two million concurrent users,
estimated from data collected on a daily basis in May, 2019~\cite{TorMetrics,
TorUsersQA}. There are more than 25K I2P relays, estimated during the same
period~\cite{i2p_metrics}. While Tor is primarily tailored for
latency-sensitive activities (e.g., web browsing) due to bandwidth
scarcity~\cite{McCoy2008}, I2P is more tolerant towards bandwidth-intensive
peer-to-peer (P2P) file sharing applications (e.g.,
BitTorrent)~\cite{Timpanaro2011}.

\section{Methodology}
\label{sec:method}

In this section, we present our approach of using the distributed network
of VPN Gate servers to conduct opportunistic censorship measurements at a global scale,
and approaches to measure the accessibility of different I2P services.

\subsection{Vantage Points}
\label{sec:VGVP}

From March 10th to April 10th, 2019, we observed 192K VGVPs from 3.5K
autonomous systems (ASes) located in 181 countries. Our measurements were
conducted in an \emph{opportunistic} fashion by immediately connecting to a
VGVP and running our tests as soon as the node is discovered. There are
currently more than 5K VGVPs available at any given time~\cite{VPNGate},
providing us with an abundance of vantage points to continuously measure from
various network locations. When many VGVPs become available at the same time,
we prioritize ones located in regions where we have not previously measured.

Due to the high churn rate of VGVPs (\sectionref{sec:vpngate}) and the rate
limit that we applied (\sectionref{sec:ethics}), we could conduct a total of 54K
measurements from 1.7K ASes in 164 countries. This coverage is still comparable to
its of other measurement platforms and enough to provide meaningful
insights.

\subsection{I2P Blocking Detection}
\label{sec:i2p_blocking_detection}

To access the I2P anonymity network, users typically go
through the following steps:

\begin{itemize}[noitemsep,nolistsep]

\item Download the router software package from the official I2P website
  (\href{https://geti2p.net}{geti2p.net}) or one of its mirror sites to set up
  an I2P client.

\item The client bootstraps into the network by fetching information about
other I2P relays from reseed servers.

\item The client can then communicate with its intended destinations via other
relays that were previously fetched.

\end{itemize}

\noindent
Based on this process, a censor can hinder access to I2P using several blocking
techniques, such as domain name blocking~\cite{lowe2007great, holdonDNS,
farnan2016cn.poisoning, china:2014:dns:anonymous, Pearce:2017:Iris,
Will:2016:Satellite, KoreaSNI}, TCP packet injection~\cite{weaver.2009.reset,
crandall2007.conceptdoppler}, and redirection to
block pages~\cite{gill.2015.worldwide}.
The design of our censorship detection techniques is thus centered around
these different blocking techniques.

\myparab{Domain Name Blocking.} From VGVPs, we issued DNS queries to both local
and open resolvers\footnote{We use public DNS resolvers, including Google's
8.8.8.8 and 8.8.4.4, Cloudflare's  1.1.1.1 and 1.0.0.1, and Cisco Umbrella
OpenDNS's 208.67.222.222 and 208.67.220.220.} to resolve the domain names of
the official I2P homepage (\href{https://geti2p.net}{geti2p.net}), its mirror
site (\href{http://i2p-projekt.de}{i2p-projekt.de}), and reseed servers.
Resolutions of these domain names are vulnerable to DNS-based blocking because
they can be seen by any on-path observers, making them an effective vector for
censors to block access to I2P as well as other undesired
content~\cite{lowe2007great, holdonDNS, farnan2016cn.poisoning,
china:2014:dns:anonymous, Pearce:2017:Iris, Will:2016:Satellite}.

By inspecting the traffic captured during these name resolutions and comparing
the returned IP addresses with the legitimate ones, we could detect if a DNS
response was tampered. More specifically, we aggregated DNS responses returned
from known uncensored locations (e.g., the U.S., Canada) to generate a
consensus list of legitimate responses, which was then used as ground truth. We
also queried an ``innocuous'' domain (\href{http://example.com}{example.com}) to
differentiate between spurious network errors (if any) and filtering events.

As DNS resolutions are a prerequisite to obtain the correct IP address(es) of a
domain name, it is often sufficient to conduct DNS-based blocking. Prior work,
therefore, has extensively looked at DNS-based blocking~\cite{Pearce:2017:Iris,
Will:2016:Satellite}. With the introduction of DNS over
HTTPS/TLS~\cite{rfc8484, rfc7858}, DNS-based blocking may no longer be an
effective filtering channel. Nevertheless, the current design of TLS also
exposes visited domain names in the Server Name Indication (SNI)
extension~\cite{rfc6066}. SNI provides a second channel for on-path observers
to monitor HTTPS-based sites, and thus it can be used to interrupt connections
to censored destinations~\cite{KoreaSNI, Frolov2019a}.

To also examine whether SNI-based blocking is being used by censors
in light of encrypted DNS traffic, we connected to the legitimate IP address of the
official I2P homepage~\footnote{Currently, only the official homepage is
served over HTTPS. Mirror sites are still served over HTTP.} over the VPN
tunnel of VGVPs, and then monitored if the connection was interrupted during the TLS
handshake.

\myparab{TCP Packet Injection.} The injection of TCP RST (reset) or FIN
(finish) packets is another common method for blocking connections to censored
websites~\cite{iclab_SP20} and services~\cite{Arun:foci18}. To observe this
filtering technique, it is desirable to capture and analyze network traffic
while establishing connections to tested destinations. While~i) crawling the
I2P homepage and its mirrors, and~ii) establishing TCP connections to the
reseed servers and five I2P relays (set up by
us---see~\sectionref{sec:ethics}), we also captured network traffic passing
through the VPN interface between our testing machines and VGVPs. The captured
network traffic was then analyzed to see if there was any injection interfering
with our connections.

\myparab{Block Pages.} Block pages are a form of overt censorship in which censors
explicitly let users know about their blocking
intention~\cite{gill.2015.worldwide}. Block pages can be delivered through
various methods.
A censor can poison the DNS resolution of censored websites to route
users to the block page. We observe this type of blocking from an
institutional network in South Korea
(see~\sectionref{sec:domain-name-based-blocking}). Another method is to
interfere with the TCP stream to redirect users to the block page, which we
observe in Oman, Qatar, and Kuwait (see~\sectionref{sec:blockpage}). As the
official I2P site did not change much during our measurement period, we
could simply compare the HTML body of the legitimate site with those fetched
over VGVPs to detect block pages. For future reference, when crawling the I2P
site and its mirror, we also captured a screenshot of any delivered block page.

\section{Ethical Considerations}
\label{sec:ethics}

As Internet censorship is often politically motivated~\cite{Nishihata:IIC2013,
Hardy:Usec14}, measurements involving volunteer-operated devices need to be
conducted in a careful manner~\cite{Sicker2007, Penney:FOCI12,
Jones:2015:ECC}. While there are some commercial VPN services that also provide
access to residential networks (e.g., Geosurf~\cite{geosurf},
Hola~\cite{hola}, Luminati~\cite{luminati}), there have been reports of
illicit behaviors by some of these VPNs~\cite{mi2019resident}, making them
inappropriate to use for academic purposes.
We instead opt to conduct our measurements
using VPN Gate's volunteer-run nodes for several reasons.

VPN Gate is an academic project and does not have any motivation to monetize
its service like commercial VPN providers~\cite{Ikram:IMC16, Khan:2018:VPN}.
To become a VPN server, the SoftEther VPN software requires an
operator to manually go through a process with repeated warning messages
about the associated risks of joining the VPN Gate
research network~\cite{VPNGate_guideline}. We therefore expect that VGVP operators
fully understand the potential issues of sharing their connection.

The VPN Gate software, as well as the infrastructure at the University of
Tsukuba, both have logging mechanisms to assist VPN operators in case of
complaints or disputes. Although log retention can be a security and privacy
risk for VPN users, these logs serve as an anti-abuse policy used by the
project to protect its volunteers. The University of Tsukuba, and the VPN Gate
project in particular, operates under Japan laws, and thus will only provide
logs if there are valid reasons to obtain them by authorized entities. Foreign
authorities who want these logs will have to adhere to Japan laws and request
them via the Minister for Foreign Affairs~\cite{vpngate_disclosure}.

Our study of the I2P anonymity network, which comprises thousands of users,
must be performed in a responsible manner that both respects user
privacy~\cite{Sicker2007, Zevenbergen2013} and ensures that our measurements
do not interfere with the normal operation of the I2P
network~\cite{I2PResearchGuidelines}. Therefore, we apply an average rate
limit of three measurements per minute to make sure that our experiments do
not saturate any I2P or VPN Gate services, thus affecting other users.

Our measurements involve connecting to other I2P relays whose IP
address(es) may be considered as sensitive information under certain
circumstances, as they could be used to identify individuals. To prevent this
privacy risk, we set up our own I2P relays for this study and only test the
connectivity between VGVPs and these relays. Setting up our own relays
provides several benefits.
First, they help to avoid any privacy risks associated with
using other relays. Second, they improve the accuracy of our
measurements, since I2P is a dynamic network in which relays join and leave the
network frequently. The high churn rate of relays may negatively affect our
observations. Finally, measurements on our own I2P relays will not interrupt
normal usage of other relays in the network. 

More importantly, we strictly adhered to the I2P community's
guidelines~\cite{I2PResearchGuidelines} for conducting studies on the I2P
network. In accordance with these guidelines, we contacted the I2P team to
discuss the purposes of our measurements. While capturing the network traffic
of our measurements, we did not capture any traffic of other I2P or VPN Gate
users. In particular, we only ``listened'' for traffic passing through the VPN
interface between our testing machines and VGVPs. This network traffic
contains only packets generated by our tests, as discussed
in~\sectionref{sec:i2p_blocking_detection}.

\section{Data Analysis}
\label{sec:data_analysis}

Between March 10th to April 10th, 2019, we conducted a total of 54K
measurements from 1.7K ASes in 164 countries, and  detected I2P blocking
activities in five countries: China, Iran, Oman, Qatar, and Kuwait. In the
following section, we discuss the different types of blocking we observed. A
summary of our findings is provided in Table~\ref{tab:sum_block} in the
Appendix.

\subsection{Domain Name Blocking}
\label{sec:domain-name-based-blocking}

\myparab{DNS-based Blocking.} China was dominant in terms of DNS-based blocking
events across all VGVPs used. Based on the method described
in~\sectionref{sec:i2p_blocking_detection}, we detected DNS poisoning attempts
when resolving domains of the I2P homepage and reseed servers. While open
resolvers are often used by Internet users to bypass local censorship, we found
that China's Great Firewall (GFW)~\cite{china-gfw, lowe2007great} also poisons
DNS responses from our selected open resolvers when resolving censored
domains. However, we could obtain the correct DNS records for the
``innocuous'' domain (i.e., \href{https://example.com}{example.com}), which
means that despite monitoring all DNS resolutions passing by, the GFW does not
block access to open resolvers and only poisons responses for censored
domains.

\begin{table*}[t]
\centering
\begin{adjustbox}{width=2\columnwidth,center}
\begin{tabular}{|c|c|c|}
\hline
Chinese ASes & Censored domains & Most abused /24 subnets\\[0.5ex] 
\hline

AS134762, AS17816, AS4134, AS4808    & geti2p.net, i2p-projekt.de(*)        & 64.33.88.0, 203.161.230.0, 31.13.72.0, 4.36.66.0,\\
AS4812, AS4837, AS56005, AS56040     & reseed.i2p-projekt.de, netdb.i2p2.no & 74.86.151.0, 74.86.12.0, 69.63.184.0, 69.171.229.0,\\
AS56041, AS56042, AS56046, AS9808(*) & i2p.mooo.com, i2p.novg.net(*)        & 66.220.152.0, 66.220.149.0, 31.13.84.0 \\
\hline

\end{tabular}

\end{adjustbox}
\vspace*{-2mm}
\caption{Censored domains in China and top IP addresses that are most
frequently abused for poisoning DNS responses.}
\vspace*{-5mm}
\label{tab:china_dns_block_Stats}
\end{table*}

Table~\ref{tab:china_dns_block_Stats} lists the ASes from which we detected
poisoned DNS responses. The second column shows censored domains. The third
column shows /24 subnets that were most frequently abused by the GFW to inject
falsified DNS responses. While Pakistan, Syria, and Iran poison DNS responses
with NXDOMAIN~\cite{rfc8020, syriacensorship, pakistancensorship} or reserved
local IP addresses~\cite{Aryan:2013}, making them easier to distinguish, China
often falsifies DNS responses with public IP addresses belonging to other
non-Chinese organizations~\cite{lowe2007great, brown2010dns, china-gfw,
farnan2016cn.poisoning, Pearce:2017:Iris}.

Of these abused IP addresses, several were observed by previous studies.
Similar to an initial observation by Lowe et al.~\cite{lowe2007great}, we
observed 64.33.88.161, 203.161.230.171, and 4.36.66.178 among the most abused
addresses. Similar to the findings of Pearce et al.~\cite{Pearce:2017:Iris}
and Farnan et al.~\cite{farnan2016cn.poisoning}, 8.7.198.45, 59.24.3.173, and
78.16.49.15 were observed, though they were not within the group of most abused
addresses. In addition to those seen by previous work, to our surprise, we
found many new abused IP addresses, most of which belong to Facebook and
SoftLayer.

Although the IP addresses that are used to poison DNS responses are similar
across most ASes, showing a centralized list of IPs that are being abused by
the GFW, the block list of domains and blocking mechanisms seem to be
implemented differently at different network locations. For instance, in
addition to four domains poisoned at most ASes in China, we
observed DNS poisoning attempts at AS9808 (Guangdong Mobile Communication)
when resolving \emph{i2p-projekt.de} and \emph{i2p.novg.net}. Analyzing
packets captured from this AS, we notice that the way poisoned responses were
crafted is different from other ASes. More specifically, while poisoned
responses at other locations contain only the falsified IP addresses shown in
Table~\ref{tab:china_dns_block_Stats}, poisoned responses at AS9808 have
an additional resource record of a loopback IP address (i.e, 127.0.0.1).
Nevertheless, this phenomenon could also happen due to implementation bugs of
the GFW, as it only occurred sporadically but not consistently during
the period of our study.
Previous work has shown that the GFW may not always function as
desired~\cite{ensafi2015analyzing}.

In conclusion, our measurements show that the I2P homepage is
censored by DNS-based blocking, while its mirror is still accessible from
most network locations in China. Of the ten reseed servers that were active
during our measurement period,
three were consistently blocked by DNS poisoning. Our
observations align with findings of earlier studies.
We previously conducted
active measurements from China to test the reachability of reseed servers and
found that some of them were still accessible~\cite{hoang:imc18}. Moreover,
our I2P metrics site~\cite{i2p_metrics} shows a
consistent number of Chinese relays
during our measurement
period. A recent study by Ververis et al.~\cite{Ververis:2019}
also shows that the I2P Android App
is still available for download from the Tencent App Store
despite the removal of many other censorship circumvention applications.

\myparab{SNI-based Blocking.} As mentioned
in~\sectionref{sec:i2p_blocking_detection}, we investigated if
censors employed SNI-based blocking together with DNS-based blocking, as these
are the two main channels where visited domains are exposed. Surprisingly, we could
successfully fetch the official I2P homepage from the network locations in
China, where the website was previously blocked by the GFW's DNS poisoning.
Although OONI recently reported that China uses SNI-based blocking
together with DNS-based blocking to censor all domains belonging to
Wikipedia~\cite{ChinaSNI}, our findings show that this technique is not fully
employed for all censored domains. In other words, the GFW may apply different
blocking techniques against different domains and services.

\myparab{Institutional Filtering and Leakage of DNS Injection.} Apart from
poisoned responses observed in China, we also detected DNS-based blocking at
AS38676, AS9848, and AS1781 in Korea. For AS1781, which is managed by the
Korea Advanced Institute of Science and Technology, poisoned DNS responses
contain only one static IP addresses (143.248.4.221). Upon visiting the
webpage hosted under this IP address, it becomes obvious that the Institute
has deployed a firewall to filter anonymity services. Note that filtering
activities observed at institutional networks should be carefully analyzed and
not characterized as national-level filtering.
Of the 1.7K networks we had access
to, there were 64 institutional networks in 17 countries. However, after
excluding VGVPs from these networks, we still had access to other VGVPs
located in residential networks in these 17 countries.

Next, we noticed that the pattern of poisoned responses in AS38676 and AS9848
was not consistent. More specifically, we only observed poisoned responses
sporadically on some days, while we could obtain correct responses on some other
days. Further investigation from the captured network traffic showed that
poisoned responses were only injected when querying the open resolvers but not
local resolvers. Therefore, it is clear that operators of these two networks
do not block access to I2P. Moreover, the set of falsified IP addresses is
similar to those observed in China, as shown in
Table~\ref{tab:china_dns_block_Stats}. This is likely the case of China's
censorship leakage because China inspects and censors both egress and ingress
network traffic passing through the GFW. Due to the geographical proximity of
Korea and China, it is likely that our DNS queries sent from Korea to open
resolvers passed through China's network, and thus got
poisoned~\cite{Anonymous:CCR2012, demchak2018china}.

\subsection{TCP Packet Injection}
\label{sec:tcp_injection}

During our measurement period, we detected injection of TCP packets while
visiting the official I2P homepage and its mirror site in four countries. More
specifically, we found that the I2P mirror site was blocked in Iran, while the
official website was still accessible over HTTPS. Analyzing the captured
network traffic, we could detect TCP packets injected immediately after
the HTTP GET request containing the hostname was sent out. The injected TCP
packets contain HTTP 403 Forbidden, thus disrupting the normal connection.

We also found injected TCP packets from VGVPs located in Oman and Qatar. These
two censors use the same blocking techniques to prevent access to both
official and mirror sites. When connecting to the HTTP mirror site, TCP
packets were injected immediately after the HTTP GET request, redirecting
users to block pages (see~\sectionref{sec:blockpage}). When connecting to the
official site (over HTTPS), SNI-based blocking was used to interrupt the
connection. More specifically, although the TCP handshake between
\href{https://geti2p.net}{geti2p.net}
and our VGVPs in these two countries could
successfully complete, immediately after the TLS client-hello message was sent
out, a TCP RST packet was then injected, terminating the TCP stream.

Similar blocking activities with Iran were also detected in Kuwait. More
specifically, the I2P homepage was still accessible, while its mirror site was
blocked by means of TCP packet injection, redirecting users to a block page
(see~\sectionref{sec:blockpage}). However, unlike Iran, Oman, and Qatar, where
we found filtering events in many network locations, we consistently observed
blocking activities only at AS47589 (Kuwait Telecommunication Company), while
all I2P services could be accessed normally from other network locations in
this country.

\subsection{Block Pages}
\label{sec:blockpage}

Although explicit block pages can be delivered to censored users through either DNS
poisoning or TCP packet injection, as discussed above, we mostly observed
block pages at a national level being delivered through TCP packet injection.
Comparing the HTML body of the legitimate official homepage and the HTML
fetched via VGVPs, we could simply pinpoint block pages returned by censors and
detect explicit block pages in Oman, Qatar, and Kuwait.

Based on the content of the delivered message on each block page (some
examples are provided in Appendix~\ref{sec:appendix}),
it is clear that blocking access to I2P is
required by the state law in each of these three countries. Note that although
we observed the same block pages in all network locations in Oman and Qatar, of
six networks in Kuwait (AS3225, AS42961, AS9155, AS6412, AS196921, and
AS47589) from which we conducted our measurements, we only detected censorship in
AS47589. The block page explicitly explains the site is restricted under
Internet services law in the State of Kuwait. This observation shows that
there is always region-to-region and ISP-to-ISP variation, thus necessitating
comprehensive measurements to be conducted from several network locations to
accurately attribute censorship (i.e., at a local or national level).

\subsection{Comparison with other Platforms}
\label{sec:comparison}

Among currently active censorship measurement platforms, OONI~\cite{ooni-paper} is
comparable to ours in terms of coverage, with about 160 countries and 2K
network locations as of 2019. ICLab~\cite{iclab_SP20} is
similar to ours in terms of censorship detection techniques and the design
decision of using VPN vantage points to measure network filtering.

OONI provides installation packages for several platforms, including Raspberry
Pi, OS X, Linux, Google Play, F-Droid, and Apple's App Store, making it easier
for testers from around the world to download and run the package. OONI,
however, does not have full control over the measurements conducted by its
volunteers. As a result, these measurements may be interrupted by unexpected
spurious network connectivity issues at the testing client side, making the
collected data unusable or even unreliable in some
cases~\cite{Yadav2018:ooni-flaw}. 

We analyzed OONI data collected during the same study period as ours to
examine if OONI detected similar blocking events. The domain name of the I2P
homepage has been on the global test list of OONI since February,
2019~\cite{clbl:i2p}. However, we could not find any OONI tests of the I2P
website conducted from the countries in which we detected I2P censorship
(\sectionref{sec:data_analysis}), except for one test conducted in Iran. Upon
closer inspection of this test attempt, conducted by an OONI volunteer in
Iran~\cite{ooni:iran:i2p}, we found that the test could not provide reliable
data due to a control failure.

We collaborated with the authors of ICLab~\cite{iclab_SP20} to use
their platform for conducting I2P censorship measurements. However, we did not
detect any filtering activities from measurement data obtained by ICLab.
Understandably, ICLab has more limited coverage of
62 countries, as of December 2018.
Among the five countries in which we detected I2P blocking events,
IClab only had vantage points in Iran and China. However, connections to them
were intermittent, thus could not provide us with reliable data. This is
one of the advantages of our proposed infrastructure compared with commercial
VPN services, as gaining access to networks in countries with less freedom
of expression can be challenging.

\section{Related Work}
\label{sec:related_work}

Many works have conducted censorship measurements in separate countries. The
GFW of China has been extensively studied due to its
significance~\cite{claytonchina, parkchina, xuchina, ensafi2015analyzing,
wright2014cn.regional, Arun:foci18}. Some other well-known censors, including
Iran~\cite{Aryan:2013, anderson2013ir.throttling},
India~\cite{Yadav2018:ooni-flaw}, Pakistan~\cite{pakistancensorship,
khattak.2014.isp}, Syria~\cite{syriacensorship},
Yemen~\cite{dalek2015information}, Egypt, and
Libya~\cite{dainotti2013eg.ly.outages}, have also been investigated.
Throughout our this paper, we examined the blocking situation of different I2P
services in many countries. In addition to those that have been studied
previously, our study discovered explicit blockage in three more countries:
Oman, Qatar, and Kuwait.

ICLab~\cite{iclab_SP20}, OONI~\cite{ooni-paper},
Quack~\cite{VanderSloot:2018:Quack}, Iris~\cite{Pearce:2017:Iris}, and
Satellite~\cite{Will:2016:Satellite} are active platforms
capable of measuring censorship at a global scale. Despite sharing a similar
goal with us, each platform has its own drawbacks which can be complemented by
our proposed measurement infrastructure. While the design of ICLab is similar
to ours, it is challenging for the platform to obtain reliable vantage points
from commercial VPN providers in some countries of interest where we have
discovered I2P blocking activities. Although OONI is widely known for its
worldwide censorship measurement activities, Yadav et al. show that this
platform can result in some inaccuracy~\cite{Yadav2018:ooni-flaw}.

Satellite-Iris~\cite{Satellite-Iris}, a combination between two prior works
(Satellite~\cite{Will:2016:Satellite} and Iris~\cite{Pearce:2017:Iris}),
uses open DNS resolvers in the IPv4 space to detect DNS-based network
filtering. With a similar design that uses Zmap~\cite{Durumeric:2013} to probe
the whole IPv4 space to detect open servers, Quack~\cite{VanderSloot:2018:Quack} scans for public echo
servers and takes advantage of these servers to measure censorship. The
primary goal of Quack is to detect censorship of websites, but not send
or receive actual HTTP(S) packets. Instead, the platform crafts packets that
mimic HTTP(S) requests, which echo servers will reflect back to the testing
client. Nonetheless, Quack's authors have acknowledged the possibility of
false negatives when the censor only looks for HTTP(S) traffic on the usual ports
(80 and 443) since the echo protocol operates over port 7~\cite{rfc862}.

\section{Discussion}
\label{sec:discussion}

We have introduced an infrastructure that can remedy the common
challenge faced by current Internet censorship measurement platforms, which is
the trade-off between depth of measurement and breadth of coverage. The
infrastructure is built on top of a network of distributed VPN servers,
providing us with not only an abundance of vantage points around the
world, but also the flexibility of the VPN technology in applying
different testing techniques to measure
network filtering activities at a global scale.

Due to the limitations discussed in~\sectionref{sec:background}, however, we
do not consider the proposed infrastructure as a replacement of existing
measurement platforms. Instead, it should be used as a complementary tool for
conducting additional measurements from locations inaccessible to current
platforms, providing more data to analyze and improve the accuracy of
censorship measurements. For example, OONI volunteers can connect to
VGVPs and run tests to increase the coverage and accuracy of OONI's data.
Similarly, ICLab could integrate VPN Gate's OpenVPN configuration files into
its measurement platform to increase the coverage of both network locations
and countries of interest.

Our findings show that the most dominant filtering technique is based on
domain names. Currently, visited domain names can be observed in two channels:
DNS queries and the SNI extension (if HTTPS is supported), making them effective
filtering vectors for on-path observers.
While DNS over HTTPS/TLS~\cite{rfc8484, rfc7858} and
ESNI~\cite{rfc-draft-ietf-tls-esni-03} are still being developed and have not
been widely adopted yet, we believe that domain name blocking will no
longer be an effective blocking strategy once these new techniques become
standardized.\footnote{Unless users are forced to use the DNS resolvers
provided by their local Internet authority,
and they cannot use any other third-party
open DNS resolvers that support DNS over HTTPS/TLS.}

Assuming a future Internet with all traffic encrypted, it is likely that
censors will switch to employing IP-based blocking. Our measurement data shows
that the official I2P homepage, its mirror site, and
reseed servers are hosted on static IP addresses. As a result, it is
trivial for a censor to block access to these services by blacklisting all
associated hosting IP addresses. In order to cope with this problem, operators
of these domain names should consider hosting them on dynamic IP address(es)
that may also host many other websites, to discourage censors from conducting
IP-based blocking due to the cost of collateral damage of blocking many
``innocuous'' co-hosted sites.

The I2P developers have foreseen a scenario in which all reseed servers get
blocked, thus preventing new relays from joining the network. They therefore
have created a function in the I2P router software for manual reseeding. Using
this function, any active I2P relay can manually extract information of a set
of its known active relays and share it with censored relays that
do not have access to any reseed servers. Under this situation, a censor who
wants to prevent local users from accessing the I2P network will have to
harvest all IP addresses of active I2P relays and block them all. While in our
previous work we
showed that this harvesting attack could be conducted at a relatively
low cost~\cite{hoang:imc18}, we did not observe any such blocking activities
while conducting connectivity tests between VGVPs and our own I2P relays.

\section{Conclusion}
\label{sec:conclusion}

Over a one-month period, we used a network of VPN servers
distributed across 164 countries to conduct 54K measurements with the goal of
investigating the blocking of I2P at a global scale. We found that several I2P services
(e.g., the homepage, its mirror site, and a subset of reseed servers) were
blocked using different filtering techniques in five countries.

China blocks access to the official I2P homepage and a part of reseed servers
by poisoning DNS resolutions. Iran interrupts connections to the mirror site by
injecting forged TCP packets containing HTTP 403 Forbidden code. SNI-based
blocking was detected when visiting the official I2P homepage over HTTPS in
Oman and Qatar, while explicit block pages were detected when visiting the
mirror site via HTTP. Block page redirection was also detected in the network
of Kuwait Telecommunication Company when visiting the I2P mirror site.
Finally, we discussed potential approaches to help I2P tackle censorship based
on the above findings.

\section*{Acknowledgments}

We would like to thank our shepherd, Masashi Crete-Nishihata, and all of the
anonymous reviewers for their thorough feedback on earlier drafts of this paper. We also thank
Arian Akhavan Niaki, Shinyoung Cho, Zachary Weinberg, Abbas Razaghpanah,
Nicolas Christin, and Phillipa Gill from the ICLab team for collaborating with
us to run measurements to test I2P accessibility on their platform, and for
constructive discussions. We also thank Vasilis Ververis, Marios Isaakidis, and Valentin
Weber for helpful conversations and early sharing of their app store
censorship study.

This research was supported in part by the Open Technology Fund under an
Information Controls Fellowship. The opinions in this paper are those of the
authors and do not necessarily reflect the opinions of the sponsor.

\bibliographystyle{plain}
\bibliography{main}

\appendix
\section{Appendix}
\label{sec:appendix}

\begin{figure}[ht]
\centering
\includegraphics[width=\columnwidth,height=0.39\textwidth]{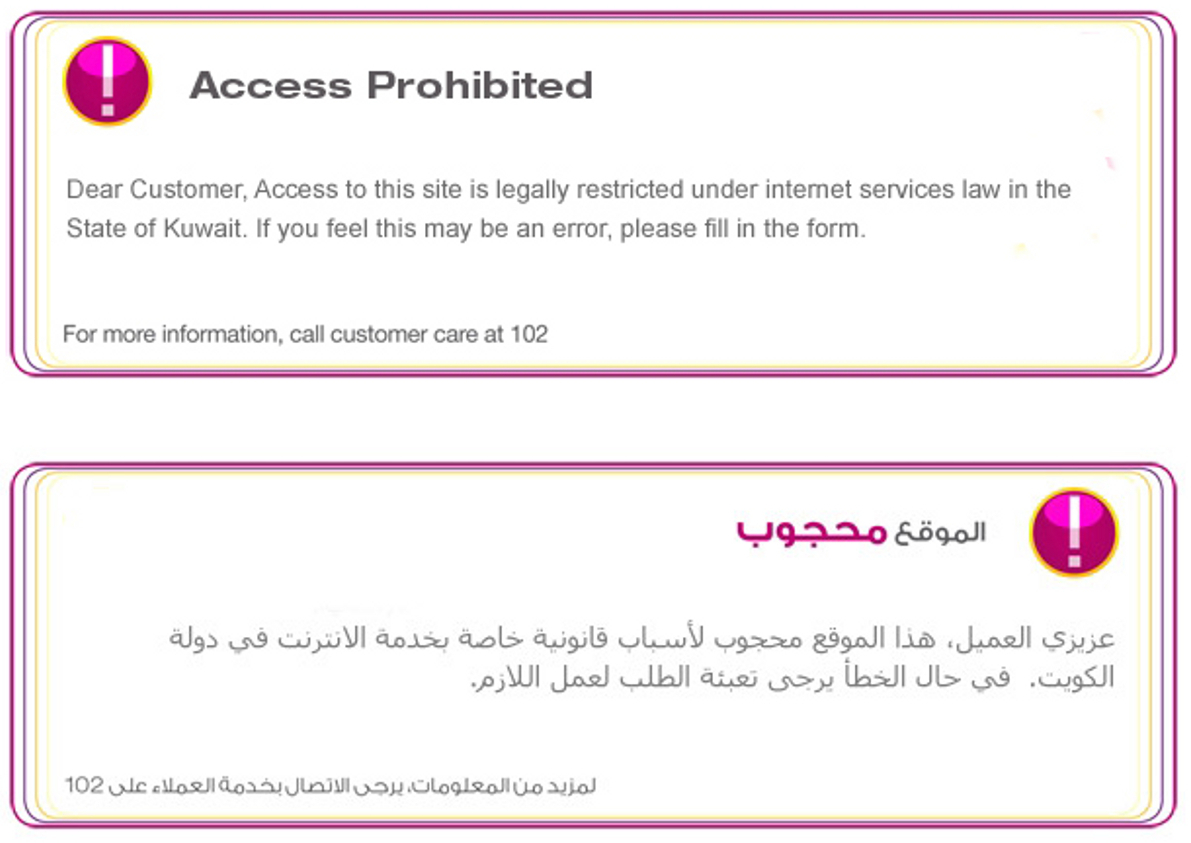}
\caption{Example block page from Kuwait.}
\label{fig:kuwait}
\end{figure}

\begin{figure}[ht]
\centering
\includegraphics[width=\columnwidth,height=0.33\textwidth]{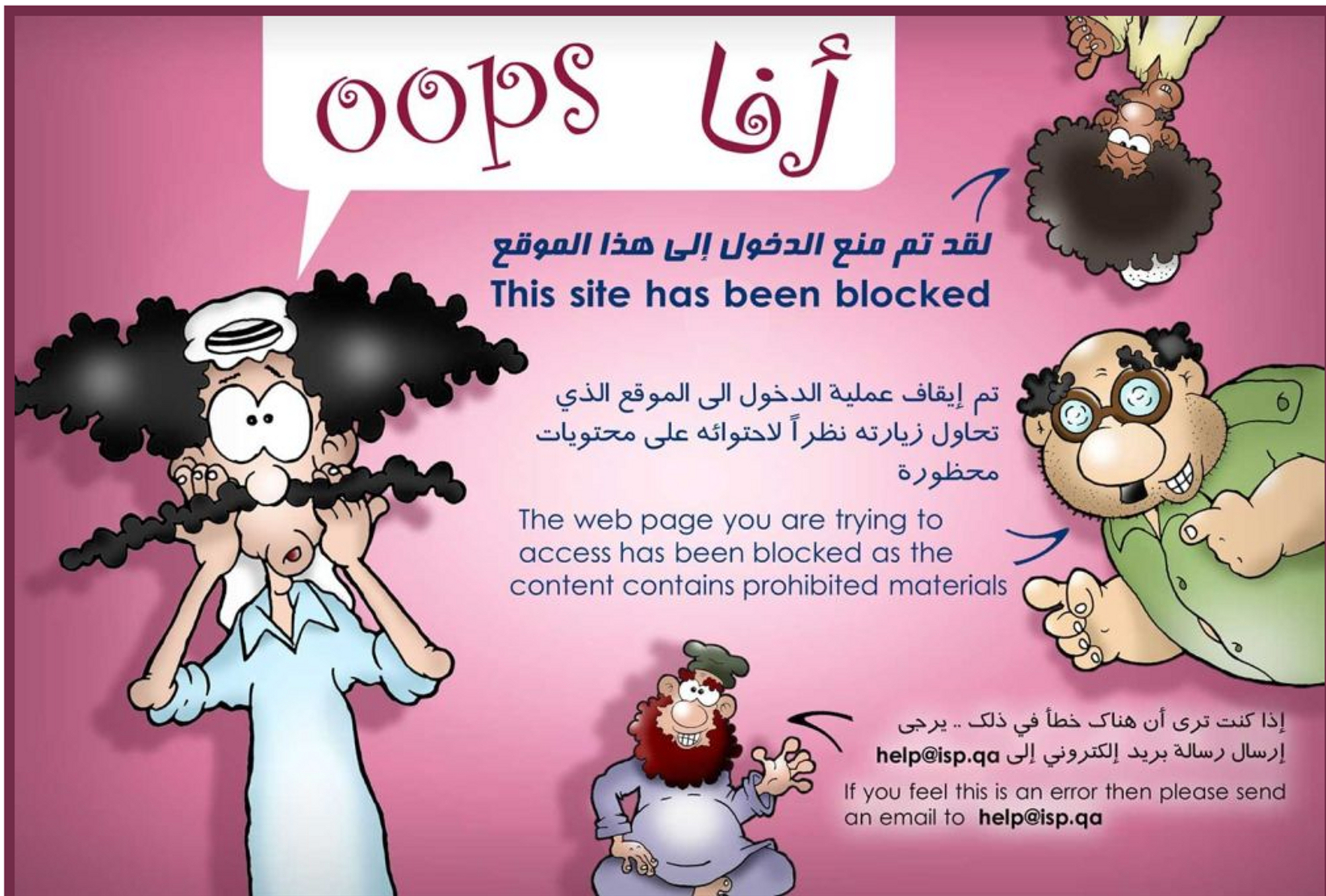}
\caption{Example block page from Qatar.}
\label{fig:qatar}
\end{figure}

\begin{figure*}[t]
\centering
\includegraphics[width=0.8\textwidth]{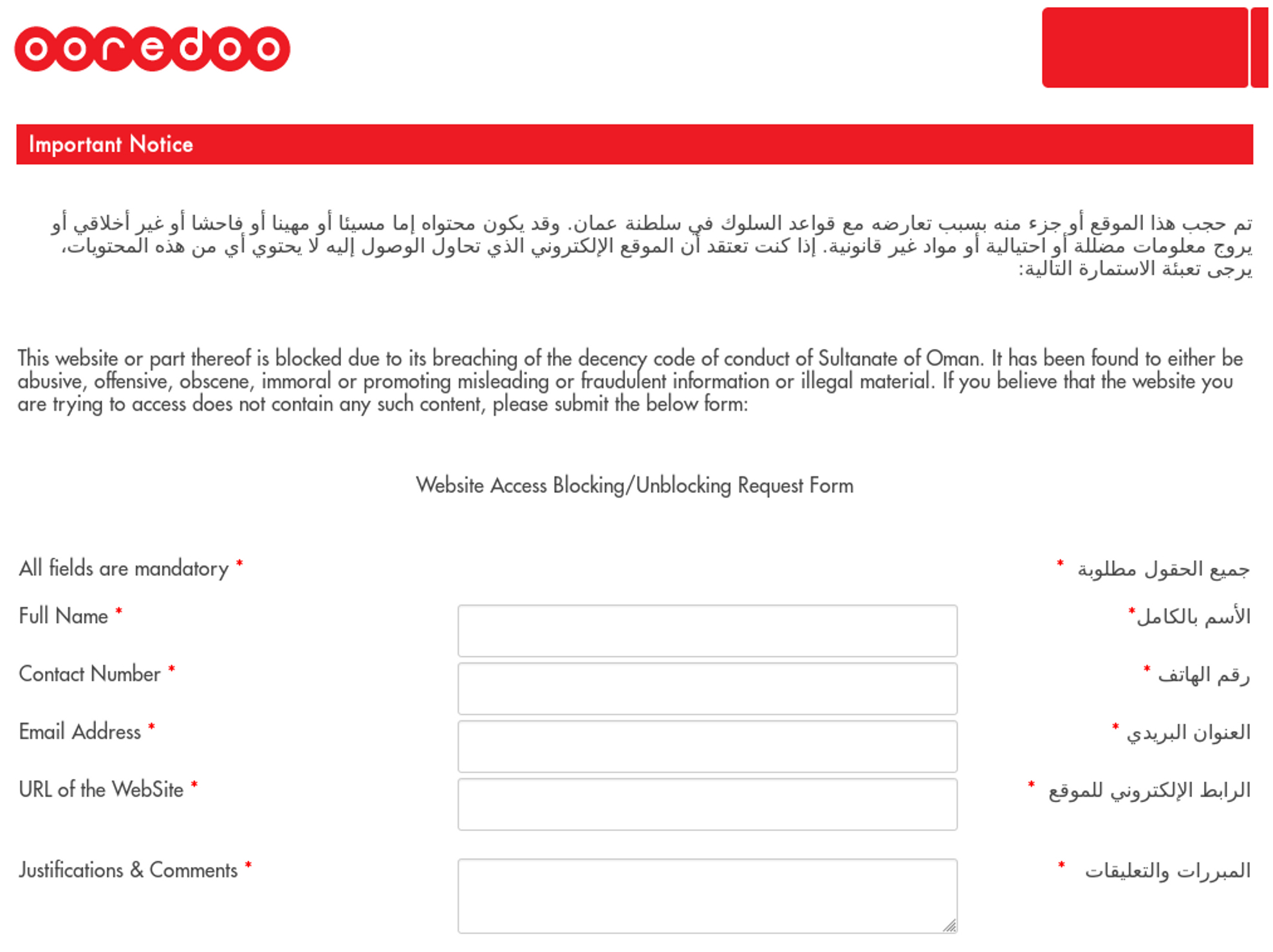}
\caption{Example block page from Oman.}
\label{fig:oman}
\end{figure*}

\renewcommand{\baselinestretch}{1.3}
\begin{table*}[t]
\centering
\begin{tabular}{|c|c|c|c|c|}
\hline
\multirow{2}{*}{Country} & \multicolumn{2}{c|}{Domain-name-based blocking}                                                                        & \multirow{2}{*}{TCP packet injection}                               & \multirow{2}{*}{Block page} \\ \cline{2-3}
                         & DNS                                                                                                       & SNI        &                                                                     &                             \\ \hline
China                    & \begin{tabular}[c]{@{}c@{}}geti2p.net\\ reseed.i2p-projekt.de\\ netdb.i2p2.no\\ i2p.mooo.com\end{tabular} & N/A        & N/A                                                                 & N/A                         \\ \hline
Iran                     & N/A                                                                                                       & N/A        & i2p-projekt.de                                                      & N/A                         \\ \hline
Oman                     & N/A                                                                                                       & geti2p.net & \begin{tabular}[c]{@{}c@{}}geti2p.net\\ i2p-projekt.de\end{tabular} & i2p-projekt.de              \\ \hline
Qatar                    & N/A                                                                                                       & geti2p.net & \begin{tabular}[c]{@{}c@{}}geti2p.net\\ i2p-projekt.de\end{tabular} & i2p-projekt.de              \\ \hline
Kuwait                   & N/A                                                                                                       & N/A        & i2p-projekt.de                                                      & i2p-projekt.de              \\ \hline
\end{tabular}
\caption{Summary of censored countries, filtered I2P services, and blocking techniques detected.}
\label{tab:sum_block}
\end{table*}

%%%%%%%%%%%%%%%%%%%%%%%%%%%%%%%%%%%%%%%%%%%%%%%%%%%%%%%%%%%%%%%%%%%%%%%%%%%%%%%%
\end{document}